\documentclass[numreferences]{kluwer}
\usepackage{epsfig}
\def\now{\number\time\ \ifcase\month\or                         
                                     January\or			
                                     February\or		%
                                     March\or			%
                                     April\or			%
                                     May\or			%
                                     June\or			%
                                     July\or			%
                                     August\or			%
                                     September\or		%
                                     October\or			%
                                     November\or		%
                                     December\fi		%
           \ \number\day,\ \number\year				%
          }							
\begin{document}
\begin{article}

\begin{opening}
\title {Effects on the structure of the universe of an\\ accelerating
expansion\thanks {Work supported in part by the US Department of Energy
(contract W-7405-ENG-36)}}
\author{George A. Baker, Jr.}
\institute{Theoretical Division, Los Alamos National Laboratory\\
University of California, Los Alamos, N. M. 87544 USA }
\date{\now}
\runningtitle{ACCELERATING EXPANSION OF THE UNIVERSE}
\runningauthor{George A. Baker, Jr.}
\begin{abstract}
Recent experimental results from supernovae Ia observations
have been interpreted to show that the rate of expansion of the
universe is increasing.  Other recent experimental results find strong
indications that the universe is ``flat.''
In this paper, I investigate some solutions of Einstein's
field equations which go smoothly between Schwarzschild's relativistic
gravitational solution near a mass concentration to the Friedmann-Lema\^\i tre
expanding universe solution. In particular, the static, {\it curved-space
extension} of the Lema\^\i tre-Schwarzschild solution in vacuum is given.
Uniqueness conditions are discussed. One of these metrics preserves the
``cosmological equation.'' We find that when the rate of expansion of
the universe is increasing, space is broken up into domains of attraction.
Outside a domain of attraction, the expansion of the universe is strong
enough to accelerate a test particle away from the domain boundary.
I give a {\it domain-size--mass relationship}. This relationship may very
well be important to our understanding of the large scale structure of the
universe.
\end{abstract}
\classification{pacs}{96.30Sf, 97.60Bw, 95.85Bh, 04.20-q}
\end{opening}
\section{INTRODUCTION AND SUMMARY}
Recently de Bernardis {\it et al.}\cite{deB} reported that a very careful
examination of the cosmic microwave background provides, as further explained
by Hu\cite{Hu}, very strong evidence that the universe is flat!

In addition,
data has recently been reported by Riess {\it et al.}\cite{Riess}, and
Perlmutter {\it et al.}\cite{Perl} on measurements of the luminosity and the
redshift of a number of high-redshift supernovae of type Ia.
The authors' best fit to their data involves a strongly curved
space-time. The curvature constant $\Omega _k$ (defined below) that they find
is about -1.05, which corresponds to a strongly curved, open universe.
However, their error bars are sufficiently large so that a flat universe
is not inconsistent with their data.  These authors have
analyzed their data on the basis of the ``cosmological equation'' between
the rate of expansion, the mean energy density, the radius of curvature of
space, and the cosmological constant. This relation has been derived
from Einstein's field equations\cite{Peb}
under the assumption of the validity of the Friedmann-Lema\^\i tre line
element at large distances.
The authors have concluded, in terms of this model
of the universe, that instead of the rate of expansion decreasing, as many
workers had thought, their data is best fit by a model in which the rate
of expansion is increasing. Riess {\it et al.}\cite{Riess} find that the
deceleration/acceleration parameter (as defined in section VI) $q_0< 0$
(acceleration) with better
than 90\% confidence, and Perlmutter {\it et al.}\cite{Perl} find the same at
the 2.86 standard deviation level of confidence.

In this paper, I investigate some different metrics which may facilitate the
investigation of some of the consequences of the reported acceleration of the
expansion of the universe. I will focus on spherically symmetric models with
a central mass concentration.

The fact that the rate of expansion of the universe is increasing leads to
the conclusion that the universe is broken up into domains of attraction.
Briefly, the underlying physics of this feature may be seen in the following
rough analysis.
The equation of purely radial motion of a test particle at rest
with respect to the Friedmann-Lema\^\i tre coordinate system is just
\begin{equation}
\ddot r =\frac{\ddot a}ar, \label{1.1}
\end{equation}
where $r$ is the proper distance, and $a$ is the universal expansion factor
which is a function of time alone.
The most simple way to approximate the effects of a gravitating mass
concentration is just to add Newton's force term so that
\begin{equation}
\ddot r =\frac{\ddot a}ar -\frac {GM}{r^2}. \label {1.2}
\end{equation}
If the expansion of the universe is decelerating, then $\ddot a < 0$, so
$\ddot r <0$ always.  Near a gravitating mass the expansion is clearly
unimportant and Newton's
laws hold with only minimal corrections.  On the other hand, if the expansion
is accelerating, then $\ddot a >0$.  Thus there will be a distance such
that the expansion of the universe exactly balances the gravitational
attraction.  Test particles at smaller distances will be accelerated
inwards.  I call this region a domain of attraction.  Test particles outside
will be accelerated outwards.  Hence in this very simple case there are
two attractors, one is the mass concentration and the other is the point
at infinity.  We give a {\it domain-size--mass relation} in equations
(\ref{8.3}) and (\ref{8.6}). In the decelerating case, there is just one
attractor and its domain of attraction is the entire space.

In the second section we remind the reader of the Lema\^\i tre-Tolman
formalism.  This formalism has the feature that the proper time appears
explicitly.  Thus space-time is divided into a set of three-dimensional,
space-like manifolds which are indexed by the proper time.  Since it is the
proper time which appears in the universal expansion function in the
Friedmann-Lema\^\i tre line element and it governs the universe as a whole,
it seems important to preserve the property that the universal expansion
factor is a function of the proper time alone.

In the third section we review three different ways to go from the
Schwarzschild metric of a planetary system to expanding space at large
distances. The McVittie solution has a problem at the Schwarzschild radius
using normal expansion factors.  The Einstein-Straus ``Swiss Cheese Model''
is unstable to perturbations and there are orbits which are discontinuous
functions of their initial conditions.  The Bona-Stela model using Liebovitz
metric insertions in a Friedmann-Lema\^\i tre background predicts an
increase in the length of the earth's year which is in strong disagreement
with observations.

In the fourth section we consider the idea that the vacuum is something
more than just empty space.  The self-energy of the vacuum may correspond
to a mass-energy density of empty space. We know on the laboratory scale
from, for example, the Casimir effect, and on the microscopic scale from
particle physics that the vacuum does have measurable effects.  So far as I
know, particle theorists have not yet been able to compute a quantitative
result for the energy density.  However, they do feel that such an idea
is extremely plausible.
We consider in this section a solution to Einstein's
field equations which has a homogeneous vacuum, mass-energy density and so
preserves the so called ``cosmological equation.''
This solution reduces precisely to
the exterior Schwarzschild solution near the central mass condensation.
In addition, it reduces to the Friedmann-Lema\^\i tre solution far from
the mass concentration.  This solution is continuously, infinitely
differentiable everywhere, except at the mass concentration.

In the fifth section, I explore some of the properties of the solution
obtained in the fourth section.  Both ``flat'' and curved space are considered.
The limiting results are as expected and the transition between the two
aforementioned limits are illustrated.  In particular the static,
{\it curved-space extension} of the Lema\^\i tre-Schwarzschild solution is given
for a mass concentration in a vacuum.

In the sixth section, I consider the implications of the increasing rate of
expansion of the universe on some of the large scale structures found in
the universe.  It is found
that this feature creates domains of attraction.  A {\it domain-size--mass
relation} is derived. Outside these domains,
the increasing rate of the expansion of the universe would, in time, be
expected to tear the structures apart.  It is suggested, from the
correspondence between the predicted size of these domains of attraction,
the size of the Local Group, and the size of the Virgo Cluster, that this effect
may well be important in any study of the large scale structure of the
universe.

\section{THE LEMA\^ ITRE-TOLMAN FORMALISM}
I will find it convenient to employ the Lema\^\i tre-Tolman\cite{Lem,Tol}
formalism.  In this case we start with the line element,
\begin{equation}
ds^2=-e^{2\alpha (\rho , \tau)}d \rho ^2 -e^{2\beta (\rho ,
\tau )}(d\theta ^2+\sin ^2\theta\, d\phi ^2) +c^2d\tau ^2, \label {2.1}
\end{equation}
where $c$ is the velocity of light, and $\tau $ is the proper time for
an observer at rest in this coordinate system.  This line element is
spherically symmetric in its structure, which is of sufficient generality
for my purposes.

The reason for our interest in this particular form of the line element
is that two important line elements are special cases.  First, the
Friedmann-Lema\^\i tre line element,
\begin{equation}
ds^2=-a^2(\tau)d \rho ^2 -a^2(\tau ) \rho ^2\left ( d\theta ^2
+\sin ^2\theta\, d\phi ^2\right )+c^2d\tau ^2 \label {2.1a}
\end{equation}
is manifestly of form (\ref{2.1}).  The second important line element
is the static Schwarzschild line element.  It is usually written in the
form,
\begin{equation}
ds^2 = -\frac {dr^2}{\displaystyle {1-\frac {2GM}{c^2r}}}-
r^2\left (d\theta ^2+\sin ^2\theta\, d\phi ^2 \right )+c^2
\left (1-\frac {2GM}{c^2r}\right )dt ^2 \label {2.1b}
\end{equation}
However, as is well known\cite{Steph} this form does not extend inside
the Schwarzschild radius $r_S=2GM/c^2$.  An alternate form has been
given by Lema\^\i tre \cite{Lem2}. It is
\begin{eqnarray}
ds^2&=&-\frac {2GM}{c^2{\mathcal R}}d \rho ^2 - {\mathcal R}^2\left (d\theta ^2
+\sin ^2\theta\, d\phi^2\right )+c^2d\tau ^2,\nonumber \\ {\mathcal R}&\equiv &
\left [{\textstyle{\frac 32}}\sqrt{2GM/c^2\, }( \rho -c\tau )\right]^{2/3}.
\label {2.1c} \end{eqnarray}
This form, without going into the full generality of the Schwarzschild
solution, has no singularity at the Schwarzschild radius.  It is a
coordinate system adapted to a freely falling observer. This metric can
easily be seen to have the form (\ref{2.1}).  As it is our goal to find
a metric which tends asymptotically to the Friedmann-Lema\^\i tre metric
at very large distances, and tends asymptotically to the Schwarzschild
metric on the scale of our planetary system, it seems appropriate
to use the Lema\^\i tre-Tolman formalism.  In this formalism,
four dimensional space-time is described by a set of three-dimensional,
space-like hypersurfaces, indexed by the time like variable $\tau $.

Next we consider the stress-energy tensor for the line element (\ref{2.1}).
First,
\begin{equation}
8\pi cT^4_1=-2\left (\beta '\dot \beta -\dot \alpha \beta '
+\dot \beta '\right ) =0, \label {2.2}
\end{equation}
where an overdot means differentiation with respect to $\tau $ and $'$
means differentiation with respect to $\rho $.  The conditions
$T^4_1 =-e^{2\alpha }T^1_4=0$ are set in order to have a time-orthogonal
coordinate system. The solution of this equation is well known to be,
\begin{equation}
e^\alpha =\frac {\beta 'e^\beta}{f(\rho)},\quad 0<f(\rho )<\infty,
\label{2.3} \end{equation}
where $f( \rho)$ is undetermined. This result allows us to eliminate
$\alpha ( \rho , \tau)$ in terms of $\beta ( \rho ,\tau )$ and
$f(\rho )$.

The other non-zero elements of the stress-energy tensor are,
\begin{eqnarray}
8\pi T^1_1&=&e^{-2\beta }-e^{-2\alpha }(\beta ')^2+2\frac{\ddot \beta}{c^2}
+3\left (\frac{\dot \beta}c\right ) ^2 -\Lambda \nonumber \\
&=&2\frac {\ddot \beta }{c^2}+3\left (\frac {\dot \beta }c\right )^2
+e^{-2\beta }[1-f^2( \rho )] -\Lambda \label {2.4} \\
8\pi T^2_2&=&8\pi T^3_3=\frac{\ddot \beta
+\dot\beta ^2 +\dot \alpha \dot \beta +\ddot
\alpha +\dot \alpha ^2}{c^2} -e^{-2\alpha }\left [\beta '' +(\beta ')^2-\alpha '
\beta '\right ] -\Lambda \nonumber \\
&=& 8\pi T^1_1+\frac {8\pi }{2\beta '}\frac {\partial T^1_1}
{\partial  \rho } \label {2.5} \\
8\pi T^4_4&=&\left (\frac {\dot \beta }c\right )^2
+2\frac{\dot \alpha \dot \beta }{c^2} +e^{-2\beta }-e^{-2\alpha }
\left [2\beta ''+3(\beta ')^2-2\alpha '\beta '\right ] -\Lambda \nonumber \\
&=&3\left (\frac {\dot \beta}c\right )^2+2\frac {\dot \beta\dot \beta '}
{c^2\beta '} + e^{-2\beta } \left [1 - f^2( \rho )
-\frac {2f( \rho )f'( \rho )}{\beta '} \right ] -\Lambda \label{2.6}
\end{eqnarray}

We see that by (\ref{2.5}), the necessary and sufficient condition that the
spatial curvature be isotropic is that $T^1_1$ be independent of $\rho $.
In addition, this condition implies that the spatial curvature is homogeneous
throughout all space.

If we replace $f( \rho )$ by unity, then by construction, $e^{\beta }
=\int e^\alpha \, d\rho $, the distance measured from the origin.  Hence
the area of a sphere is just $4\pi $ times the square of the radius, as
is given by the standard Euclidean formula. I will call this special case
the case of ``flat'' space. In the special case of ``flat'' space and
isotropic spatial curvature, (\ref {2.4}) and (\ref {2.6}) become
\begin{eqnarray}
8\pi T^1_1=8\pi T^2_2=8\pi T^3_3&=&2\frac{\ddot\beta }{c^2}
+3\left(\frac {\dot \beta }c\right )^2 -\Lambda , \label {2.8} \\
8\pi T^4_4&=& \frac {\dot \beta }c\left (3\frac {\dot \beta }c +2\frac
{\dot \beta '} {c\beta '}\right ) -\Lambda .\label {2.9}
\end{eqnarray}

We observe, that (\ref{2.4}) and (\ref{2.6}) may be rewritten as,
\begin{eqnarray}
\frac {\partial }{\partial \tau }\left \{e^{3\beta}\left [
\left (\frac {\dot \beta }c\right )^2
-\frac \Lambda 3\right ]+e^\beta [1-f^2(\rho )]\right \} &=& 8\pi T^1_1\dot
\beta e^{3\beta } \label {2.10} \\
\frac {\partial }{\partial \rho }\left \{ e^{3\beta}\left [
\left ( \frac{\dot \beta }c\right )^2
-\frac \Lambda 3\right ]+e^\beta [1-f^2(\rho )]\right \} &=& 8\pi T^4_4
\beta 'e^{3\beta }. \label {2.11}
\end{eqnarray}
The quantity
\begin{equation}
m=\frac {c^2}{2G}\left \{ e^{3\beta}\left [\left (\frac{\dot \beta }c\right )^2
-\frac \Lambda 3 \right ]+e^\beta [1-f^2(\rho )]\right \} \label {2.12}
\end{equation}
is easily recognized\cite{Lem,Kra} to be the mass equivalent to the total
energy contained within the comoving shell with radial coordinate $\rho $
at time $\tau $.

We will also be concerned with the dynamics as seen in this formalism.
The main interest will be to assess the difference between the
effects on planetary systems of the metrics which we will study
herein, and the Schwarzschild metric. In particular I am interested in
the solar system. The standard formula for the equations of motion of
a test particle\cite{Peb} is
\begin{equation}
\left (\frac {ds}{d\tau }\right )\frac d{d\tau }\left [ g_{k i}
\left (\frac {ds}{d\tau }\right )^{-1}\dot x^i\right ]=\frac 12
g_{ij,k }\frac {dx^i}{d\tau }\frac {dx^j}{d\tau } \label {4.3}
\end{equation}
where of course $dx^4/d\tau =1$.  The fourth component of (\ref{4.3}) gives
the convenient equation,
\begin{equation}
{\mathcal B}=\left (\frac {ds}{d\tau }\right )\frac d{d\tau }\left [
\left (\frac {ds}{d\tau }\right )^{-1}\right ]= -\left \{\frac {\dot \alpha
e^{2\alpha }\dot \rho ^2+\dot \beta e^{2\beta }\left [\dot \theta ^2+\sin ^2
\theta \dot \phi ^2 \right ]}{c^2}\right \} \label {4.4}
\end{equation}

Explicitly, the three equations of motion are,
\begin{eqnarray}
&&{\mathcal B}e^{2\alpha }\dot \rho +\frac d{d\tau }\left [ e^{2\alpha }
\dot \rho \right ]=\alpha 'e^{2\alpha }\dot \rho ^2+\beta 'e^{2\beta }
\left [\dot \theta ^2+\sin ^2\theta\, \dot \phi ^2\right ], \label {4.5} \\
&&{\mathcal B}e^{2\beta }\dot \theta +\frac d{d\tau }\left [ e^{2\beta }
\dot \theta \right ]= e^{2\beta }\sin \theta \cos \theta \, \dot \phi ^2 ,
\label{4.6} \\
&&\left (\frac {ds}{d\tau }\right )\frac d{d\tau }\left [e^{2\beta }\sin ^2
\theta \left (\frac {ds}{d\tau }\right )^{-1}\dot \phi \right ]=0 .
\label {4.7} \end{eqnarray}
Following the completely standard {\it modus operandi}, we make
an immediate simplification by setting $\theta =\pi /2$. Now
equation (\ref {4.6}) is explicitly satisfied and the other two equation
are simplified.  The first integral  of (\ref {4.7}) gives us the conservation
of angular momentum, to wit,
\begin{equation}
e^{2\beta } \dot \phi =C_0\left (\frac {ds}{d\tau }\right ),\label {4.8}
\end{equation}
where the $C_i$ are constants of integration.
By means of (\ref{2.1}) we may write (\ref{4.8}) as,
\begin{equation}
\dot \phi ^2=C_0^2e^{-4\beta }\left (c^2-e^{2\alpha }\dot \rho ^2-
e^{2\beta }\dot\phi ^2\right ) = \frac {C_0^2e^{-4\beta }\left (c^2-
e^{2\alpha }\dot \rho ^2\right )}{1+C_0^2e^{-2\beta }} . \label {4.8a}
\end{equation}

These steps leave us with the single equation (\ref{4.5}) for $\rho (\tau )$
to deal with. It now becomes
\begin{equation}
{\mathcal B}\dot \rho +(2\dot \alpha +\alpha '\dot \rho )\dot \rho +\ddot \rho
=\frac {\dot \phi ^2}{\beta ' }= \frac {C_0^2e^{-4\beta }\left (c^2-
e^{2\alpha }\dot \rho ^2\right )}{\beta '\left (1+C_0^2e^{-2\beta }\right )}
 , \label {4.9}
\end{equation}
where $\mathcal B$ (\ref{4.4}) may be re-expressed as,
\begin{equation}
{\mathcal B} = -\dot \alpha e^{2\alpha }\left (\frac {\dot \rho }c\right ) ^2
- \frac {C_0^2\dot \beta e^{-2\beta }\left (c^2- e^{2\alpha }\dot \rho ^2
\right )} {c^2\left (1+C_0^2e^{-2\beta }\right )} \label {4.10}
\end{equation}
Thus, (\ref {4.5}) becomes,
\begin{equation}
\ddot \rho =-(2\dot \alpha +\alpha '\dot \rho )\dot \rho +
\frac {C_0^2e^{-2\beta }\left (c^2e^{-2\beta }+\dot \beta \beta '\dot \rho
\right ) \left (c^2- e^{2\alpha }\dot \rho ^2\right )}
{c^2\beta '\left (1+C_0^2e^{-2\beta }\right )}
+\dot \rho \dot \alpha e^{2\alpha }\left (\frac {\dot \rho }c\right ) ^2
\label {4.11} \end{equation}
which gives an explicit, ordinary, non-linear differential equation for
$\rho (\tau )$.  Coupled with $\theta = \pi /2$ and equation (\ref{4.8a}),
we have the equations of motion of the test particle.

A more transparent form follows for the purpose of identifying the corrections
to the Schwarzschild dynamics results if we make the change of variables,
\begin{equation}
r=\exp [\beta (\rho ,\tau)],\quad \Rightarrow \quad \rho =\rho (r, \tau ).
\label{2.26} \end{equation}
Then we have
\begin{equation}
\dot r =\left (\dot \beta +\beta '\dot \rho \right )r. \label {2.27}
\end{equation}
Equation (\ref{4.5}) becomes,
\begin{eqnarray}
\ddot r&=&\left [\ddot \beta +\dot \beta ^2+\frac {f'}{f\beta '}\left (
\frac {\dot r}r-\dot \beta \right )^2 \right ]r +\frac {c^2C_0^2}{r^3}
+\left (\frac {\dot \beta '+\dot \beta \beta '}{c^2f^2\beta '}\right )
\left (\dot r -r\dot \beta \right )^3 \nonumber \\
&&+\frac {C_0^2\left [{\displaystyle \frac {\dot \beta }r(\dot r -\dot \beta r)
-\left (\frac {\dot r -\dot \beta r}{rf}\right )^2-\frac {\dot \beta (\dot r
-\dot \beta r)^3}{rc^2f^2}-\frac {C_0^2c^2}{r^4} }\right ]}
{r\left (1+C_0^2/r^2\right )} .\label {2.28}
\end{eqnarray}

\section{CLASSICAL MODELS}

Currently there are two general relativistic descriptions of spacetime
in popular use.  For planetary systems and other gravitationally
bound structures which are small on  the scale of the universe,
there is a static description of the behavior of spacetime. On the other
hand, for large-scale behavior, there is a time dependent description
which is appropriate as a description of phenomena such as the observed
red-shift of distant galaxies.

The classic question is, ``How can these two disparate
descriptions of spacetime possibly be reconciled with each other?'' The
first answer to this problem was given by McVittie\cite{McV}.  Using the
line element form,
\begin{equation}
ds^2=-e^\mu (d\rho ^2+\rho ^2d\theta ^2+\rho ^2\sin ^2\theta
d\phi ^2)+e^\nu dt^2 \label{9.1}
\end{equation}
He proposed the solution,
\begin{eqnarray}
e^\mu &=&{{a(t)^2}\over {[1+(a(t)\rho /2a(t)R)^2]^2}}\left (1+{m\over {2a(t)
\rho}} \right )^4  \nonumber \\
e^\nu &=& c^2\left ({{1-m/2a(t)\rho}\over {1+m/2a(t)\rho }}\right )^2 \qquad
m\equiv {{GM}\over {c^2}}\left [1+\left ({\rho\over {2R}}\right )^2\right ]
^{1/2}  \label{9.2}
\end{eqnarray}
where $G$ is Newton's constant of gravitation. Some of the virtues of this
model can be seen by examining the corresponding stress-energy tensor,
\begin{eqnarray}
8\pi T^1_1&=& {{1+(m/2a(t)\rho )^2}\over {a(t)^2R^2(1+m/2a\rho )^5
(1-m/2a(t)\rho )}} \nonumber \\
&&+{{2\ddot aa[1+m/2a(t)\rho ]+\dot a^2[1 -5m/2a(t)\rho ]
}\over{a(t)^2c^2(1-m/2a(t)\rho )}} -\Lambda \nonumber \\
8\pi T^2_2 &=&8\pi T^3_3= 8\pi T^1_1 \nonumber \\
8\pi T_4^4&=& {3\over {a(t)^2R^2(1+ m/2a(t)\rho )^5}}
+{{3\dot a^2}\over {c^2a^2}}-\Lambda .  \label {9.3}
\end{eqnarray}
In the limit as $m\to 0$ or $\rho\to \infty $ these tensor elements reduce to
those corresponding to (\ref{2.1a}),
and also in the limit as $R\to \infty $ and $\dot a\to 0$ they vanish as with
the Schwarzschild metric.  For this metric the spatial curvature is isotropic,
but not homogeneous.  The dynamical behavior of a test particle can be
computed from (\ref{4.3}).  In the usual case $\theta =\pi /2$, the equation
of motion, analogous to (\ref{2.28}), becomes,
\begin{eqnarray}
&&\ddot r -2{{\dot a}\over a}\dot r +r \left [2\left ({{\dot a}
\over a}\right )^2-{{\ddot a }\over a}\right ]+\dot \mu\left (\dot r
-{{\dot a}\over a}r \right ) \nonumber \\
&=& {1\over 2}\mu 'a\left ({{\dot r }
\over a}-{{\dot a}\over {a^2}}r \right )^2+{1\over 2}\left (
\mu '{{r }\over a}
+2\right ) \left [{{A^2a^4{e}^{-2\mu}}\over {r ^3}}\right ]
\left ({{{d} s} \over {{d} t}}\right )^2 \label{9.4} \\
&&-{1\over 2}\nu 'a {e}^{\nu -\mu}
-{1\over 2}\left ({{{d} s}\over {{d} t}}\right )^2{{d}\over {{d} t}}
\left [\left ({{{d} s}\over {{d} t}} \right )^{-2}\right ]\left (\dot r
-{{\dot a}\over a}r \right ) , \nonumber
\end{eqnarray}
where, in concert with (\ref{4.8}), $r^2e^\mu \dot \phi=A(ds/dt)$, and the
change of variables $r=a(t)\rho $ has been used.  The slow speed limit of
this equation of motion is,
\begin{equation}
\ddot r=\frac{\ddot a}ar+\frac {A^2c^2}{r^3}-\frac {GM}{r^2} \label {9.5}
\end{equation}
This equation differs from Newton's by the $\ddot a$ term.  This term is
proportion to $H_0^2$ where $H_0\approx 1.62\times 10^{-18}h_{50}$ is
Hubble's constant.  $h_{50}=1$ if $H_0=50$ km/sec/Megaparsec.  Thus in
McVittie's model, the effects on the solar system are too small
to be currently measured.  For systems like the solar system, the McVittie model
works very well. However for other systems, if we make the popular assumption
that $a(t)\propto t^\gamma $ then, as in McVittie's solution $a(t)$ is a
function of coordinate time and not of proper time, there will be a
singularity at the the Schwarzschild radius as $a(t)$ will diverge there
since $t\to \infty $.

A second answer to our question was given by Einstein and Straus \cite{ES}.
It is the so called ``Swiss Cheese Model.''  They showed that the static
Schwarzschild metric (\ref{2.1b}) can be matched to any
Friedmann-Lema\^\i tre-Robertson-Walker metric on a spherical surface
of an expanding radius in the Schwarzschild metric, but of a
constant radius in the Friedmann-Lema\^\i tre-Robertson-Walker metric.
It has been found
however\cite{Sato,LP} to be unstable with respect to
perturbations. In addition, there is another unphysical aspect of this
model.  The slow-speed dynamical equations in the interior Schwarzschild
($\vec r_{\rm i}$)
and in the exterior Friedmann-Lema\^\i tre ($\vec r_{\rm e}$) regions are,
respectively,
\begin{equation}
\frac {d^2\vec r_{\rm i}}{dt^2}=-\frac {GM}{r^3_{\rm i}}\vec r_{\rm i},\qquad
\frac {d^2\vec r_{\rm e}}{dt^2}=\frac {\ddot a}a\vec r_{\rm e} .\label {9.6}
\end{equation}
For ease of exposition, we choose $a(t)=(t/t_0)^{2/3}$.  Thus the general
solution in the exterior region is $\vec r_{\rm e}=\vec A t^\gamma +\vec B
t^{1-\gamma}=\vec A t^{2/3}+\vec B t^{1/3}$, which is the equation of a
parabola.

As an illustration of the behavior of the ``Swiss cheese model''
I have computed the following trajectories.  I use as
a unit of time the Hubble time, that is $1/H_0$.
As a unit of distance I use $\root 3 \of{M_\odot G/H_0^2}$
which is about $25h_{50}^{-2/3}$ million Astronomical units. $M_\odot $ is the
mass of the Sun.  I set $t_0=1$.  The metric interface
is at $r=t^{2/3}$ in these units for flat spacetime, as mentioned above.
In order to follow a Friedmann-Lema\^\i tre trajectory the test particle's
distance from the origin must be larger at every time than that for the
interface. Let us take in rectangular coordinates the initial conditions,
$x=\lambda ,\;\dot x =0,\; y=0,\; \dot y=\lambda $. The trajectory in
the external region is
\begin{eqnarray}
x&=&\lambda \left [-\left ({t\over {t_0}}\right )^{2/3}+2\left ({t\over {t_0}}
\right )^{1/3}\right ] \nonumber \\
 y&=&3t_0\lambda \left [\left ({t\over {t_0}}\right )^{2/3}-
\left ({t\over {t_0}}\right )^{1/3}\right ] . \label{9.6a}
\end{eqnarray}
Thus,
\begin{equation}
\lambda ^2\left [10t^{4/3}-22t +13t^{2/3}\right ] > t^{4/3} .\label{9.7}
\end{equation}
The trajectory will intersect the interface if equation (\ref{9.7}) is an
equality.
By means of the quadratic formula, an intersection will occur if
\begin{equation}
t^{1/3}={{11\pm\sqrt{-9 +13/\lambda ^2}}\over {10 -1/\lambda ^2}}.
\label{9.8}
\end{equation}
It will be observed that for $\lambda < \sqrt{13}/3 $ there are two
real roots.  If $\lambda =1$, then $t^{1/3}=1,\; 13/9$.  On the other
hand, if $\lambda > \sqrt{13}/3$, the roots are imaginary, so there are
no intersections. If $\lambda =\sqrt{13}/3$ there is a double root at
$t^{1/3}=13/11$.  In this case the parabolic trajectory just grazes the
metric interface.

In figure 1, I illustrate the two different trajectories when $r_0=
\sqrt{13}/3$ and $\dot \phi _0=1.0$. In the case where $r_0$ is just
any arbitrary amount smaller, the expanding interface overtakes the
test particle following its Friedmann-Lema\^\i tre parabolic trajectory
and it must then follow the static Schwarzschild equations of motion.
The Schwarzschild metric takes over at $t=(13/11)^3$ as explained above,
and after that the trajectory is an ellipse with
semimajor axis 7.6287 and the semiminor axis 3.986.  These imply an
eccentricity of 0.690\thinspace 68 and a semilatus rectum parameter of
$p= 2.086$. On the other hand, if $r_0$ is any arbitrary amount larger,
it escapes the moving interface and continues to follow the parabolic
trajectory. It is evident from figure 1 that future trajectories are,
in some cases, discontinuous functions of the initial conditions for the
``Swiss cheese model.''

Put another way, the Schwarzschild metric permits closed orbits and
the Friedmann-Lema\^\i tre metric does not.  In terms of the latter
coordinates, one can choose initial conditions so that the parabolic
trajectory just grazes the metric interface (fixed radial coordinate
in this metric) and the speed is low enough that for infinitesimally
different initial conditions the trajectory crosses the interface and
is caught in a bound state, or alternatively misses the interface and
proceeds on its parabolic trajectory.  All these effects take place in
supposedly empty space of the order of $25h_{50}^{-2/3}$ million AU from
a mass concentration of size $M_\odot $, and are quite counter to one's
physical intuition that such discontinuities should not occur there,
except perhaps if the central void was created by an explosion leaving a
remnant star behind.

A third answer was provided by Bona and Stela\cite{BS}. It is a different
type of ``Swiss Cheese Model.''  They insert in a flat space,
Friedmann-Lema\^\i tre-Robertson-Walker background a spherical patch in which,
instead of the Schwarzschild metric, the Liebovitz\cite{Lie} solution is
inserted. In the classification of Krasi\'nski\cite{Kra}, this solution is
in the ``$\beta ' \neq 0$'' subfamily of the Szekeres-Szafron family of
solutions.  Briefly put, this solution results from setting $T_1^1$ equal to
its Friedmann-Lema\^\i tre value, {\it i.e.}, by (\ref{2.4}),
\begin{equation}
8\pi T^1_1=2\frac {\ddot \beta }{c^2}+3\left (\frac {\dot \beta }c\right )^2
 -\Lambda = 2\frac{\ddot a}{ac^2}+\left (
\frac {\dot a}{ac}\right )^2 -\Lambda \label {9.9}
\end{equation}
Since the right hand side of (\ref{9.9}) is independent of $\rho $, the
spatial curvature is isotropic by (\ref{2.5}), and is homogeneous.

As (\ref{9.9}) is a generalized Riccati equation for $\beta $, it can be
integrated. The general solution is,
\begin{equation}
e^{\beta (\rho ,\tau )}= a(\tau )\left [C_1(\rho )+C_2(\rho )w(\tau )\right ]
^{2/3},\quad
{\rm where}\quad w(\tau )= \int _{\tau _0}^\tau \frac {dt} {a(t)^3}.
\label {9.10}
 \end{equation}
For this solution, we can compute that,
\begin{equation}
8\pi T^4_4=3\left \{\frac {\dot a}a+\frac {2C_2(\rho)\dot w}
{3[C_1(\rho )+C_2(\rho )w]}\right \}\left \{ \frac {\dot a}a +\frac {2C_2'
(\rho )\dot w}{3[C_1'(\rho )+C_2'(\rho )w]}\right \} \label {9.11}
\end{equation}

The dynamics of this model reveal that its predictions are at considerable
variance with the observations of planetary astronomy. The slow-speed limit of
the equations of motion follows directly from (\ref{2.28}) and is,
when we remember that $C_0=O(c^{-1})$ in the slow speed limit,
\begin{equation}
\ddot r=\frac {\ddot a}a r-\frac {2C_2(\rho )^2}{9a^3r^2}-\frac {2 \dot a
C_2(\rho )r}{3a(ra)^{3/2}}+\frac {c^2C^2_0}{r^3}, \label {9.12}
\end{equation}
where $r(\rho ,\tau )=e^{\beta (\rho , \tau )}$.
In order to agree with Newton's result, we must choose $C_2(\rho )=
-\frac 32\sqrt{2GM}$. The choice of $C_1(\rho )$ does not affect the dynamics,
so we pick $C_1=\frac 32\sqrt{2GM/c^2}\rho $. This choice suffices to cause
$T^4_4\to 3(\dot a/a)^2$ as $\rho \to \infty$. (Note that different choices
of $C_1$ and $C_2$ are required for the zero mass case.)

This result is substantially unique for a model which has isotropic spatial
curvature, a non-zero central mass concentration, and is asymptotic, at large
distances from that mass, to the Friedmann-Lema\^\i tre, expanding universe
model. However, this model leads, by (\ref{9.12}), to the results,
\begin{equation}
\ddot r =\frac {\ddot a}ar+ \frac {\dot a}{a^2}\sqrt {\frac {2GM}{ar}}
-\frac {GM}{a^3r^2}+\frac {c^2C_0^2}{r^3}. \label{9.13}
\end{equation}
The problem to be noticed is the factor of $a^3$ in the denominator of
the Newton gravitational attraction term.  The import of this factor is
that the radius of the earth's orbit will be proportional to $a^3$ as the
gravitational term and the angular momentum term dominate the other terms.  By
Kepler's law that the ratio of the squares of the periods is proportional
to the cube of the diameters, the year will be proportional to $a^{9/2}$
This means that the year should be increasing at the rate of about
$0.73h_{50}$ seconds per century. Here $h_{50}$ is the Hubble constant in
units of 50 kilometers per second per Megaparsec.  Since the value currently
accepted by the International Astronomical Union is $+0.0095$ seconds per
Julian century\cite{Jac}, the predictions of this model are in very serious
disagreement with observations. The above results are in terms of the
time variable $\tau $.  Properly done, we need the results in terms of
time measured on the Earth for the Liebovitz line element.  The re-expression
of this result in terms of ``Earth time'' involves a correction of the order
of $H_0 GM_\odot /(c^2R_\oplus )$ where $M_\odot $ is the mass of the
Sun and $r_\oplus $ is the radius of the orbit of the Earth.  This quantity
is of the order of $10^{-8}$ smaller than $H_0$ which is of the order of
the predicted effect.  Hence, this latter correction is negligible in this
case. There is, of course, also a time independent correction to the
relation between earth time and $\tau $ of the order of $GM_\odot /(c^2
R_\oplus )$ but it is not relevant to the effect discussed here.

The conclusion of this section is that in one way or another, all the models
reviewed that join the Schwarzschild metric at small distances and the
expanding universe metric at large distances have problems of one sort or
another. Panek \cite{Pan} has stress the need to go beyond the ``Swiss
Cheese Model'' and has analyzed numerically several different
density profiles.  In the next section, we will analyze the effect
of a mass concentration inserted in a universe with a homogeneous
mass-energy density.

\section{HOMOGENEOUS, MASS-ENERGY DENSITY SOLUTIONS}

In line with the idea that the vacuum has a homogeneous self-energy, we
consider homogeneous mass-energy density solutions to Einstein's field
equations. Our case of interest is the embedding of a mass concentration in
expanding, curved space.  We impose the condition,
\begin{eqnarray}
8\pi T_4^4 &=&3\left (\frac {\dot \beta}c\right )^2+2\frac {\dot \beta
\dot \beta '} {c^2\beta '} + e^{-2\beta } \left [1 - f^2( \rho )
-\frac {2f( \rho )f'( \rho )}{\beta '} \right ] -\Lambda \nonumber \\
&=& 3\left ( \frac {\dot a}{ca}\right )^2+\frac 3{(aR_0)^2} -\Lambda
\label{6.11} \end{eqnarray}
which is the Friedmann-Lema\^\i tre value in curved space.  Since the
right hand side is a function of $\tau $ alone, we may integrate
this equation, as at (\ref{2.11}), with respect to $\rho $ to give,
\begin{equation}
\dot \beta ^2+c^2e^{-2\beta }\left [ 1-f^2(\rho )\right ]=
\left (\frac {\dot a}a\right )^2+\frac {c^2}{(aR_0)^2}+ C_3(\tau )e^{-3\beta }.
\label {6.12}\end{equation}
At short distances, we wish to match the Lema\^\i tre-Schwarzschild metric
(\ref{2.1c}).  We find, upon substituting in (\ref{6.12}), for $\dot \beta $
that
\begin{equation}
2GMe^{-3\beta } +c^2e^{-2\beta }\left [ 1-f^2(\rho )\right ]\approx
\left (\frac {\dot a}a\right )^2+\frac {c^2}{(aR_0)^2}+ C_3(\tau )e^{-3\beta }.
\label {6.2a}\end{equation}
Thus, we shall choose $C_3(\tau )=2GM$ independent of $\tau $. For later
convenience we rewrite (\ref{6.12}) as
\begin{equation}
\dot \beta ^2= \left (\frac {\dot a}a\right )^2+\frac {c^2}{(aR_0)^2}+
2GMe^{-3\beta } -c^2e^{-2\beta }\left [ 1-f^2(\rho )\right ] .\label {6.13}
\end{equation}
If we are given $\beta (\rho _0,\tau _0)$, then this non-linear, first-order
differential equation may be integrated with respect to $\tau $ to give
$\beta (\rho _0, \tau )\; \forall \;\tau $.

In the special case $M=0$ we have the Friedmann-Lema\^\i tre solution,
and the special case, $\dot a=\ddot a=0,\; R_0=\infty $ gives us agreement
with the Lema\^\i tre form of the static Schwarzschild metric (\ref{2.1c}).

Next we compute $T^1_1$ from (\ref{2.4}) and the differentiation of
(\ref{6.13}).  We obtain,
\begin{eqnarray}
&&8\pi c^2T^1_1=3\left (\frac {\dot a}a\right )^2+\frac {3c^2}{(aR_0)^2}
\label {6.14} \\
&&+\frac {\displaystyle 2\left [\frac {\dot a\ddot a}{a^2}-\left (\frac {\dot a}
a\right )^3-\frac {\dot ac^2}{a^3R_0^2}\right ]}{\displaystyle \left \{
\left ( \frac {\dot a}a\right )^2+\frac {c^2}{(aR_0)^2}+2GMe^{-3\beta }
-c^2e^{-2\beta}\left [1-f^2(\rho )\right ] \right \}^{1/2}} \nonumber
\end{eqnarray}
which reduces correctly in the two special cases ``flat space'' with
$\dot a=\ddot a=0$, and
$M=0$.  Since $T^1_1$ is not independent of $\rho $, the spatial curvature is
neither homogeneous nor isotropic.  It varies,
as the distance from the mass concentration increases, from $3(\dot a/a)^2
+ 3c^2/(aR_0)^2$ according to (\ref{6.14}),
to the Friedmann-Lema\^\i tre limit $2(\ddot a/a)+(\dot a/a)^2+c^2/(aR_0)^2$
as the distance tends to infinity.  The McVittie solution mentioned above has
isotropic, but inhomogeneous, spatial curvature. The difference with
this solution lies in the choice here that the universal expansion factor
is a function of the proper time and his choice that it is a function
of his coordinate time, which is not an invariant quantity.

So far, we have not had to discuss $\beta '$ or $\dot \beta '$, as $\dot \beta $
given by (\ref{6.13}) has sufficed for all our computations. The general
equations do not determine the behavior of $\beta $ as a function of $\rho $.

We observe that in the case of the expanding, curved-space of the
Friedmann-Lema\^\i tre model universe, that
\begin{equation}
\frac {c^2}{(aR_0)^2}-c^2e^{-2\beta}\left [1-f^2(\rho )\right ]=0\label{6.15}
\end{equation}
Since we wish our solution to tend asymptotically to this limit,
we choose to require this result in the long distance limit.
As a result of this consideration, we choose as initial conditions for
(\ref{6.13})
\begin{eqnarray}
\exp \beta (\rho , \tau _0) &=& R_0\sin \left (\frac \rho {R_0}\right ), \label
{6.16} \\
f(\rho )&=&\sqrt {1-R_0^2\sin ^2\left (\frac {\rho }
{R_0}\right )/R_0^2}, \label {6.17}
\end{eqnarray}
where, as the dependence on $R_0$ is actually a dependence on $R_0^2$, there
is no problem with the case where $R_0^2<0$. The continuation is just
$x\sin (\rho/x)\mapsto |x|\sinh (\rho/|x|)$.  Using these initial conditions
and $a(\tau _0)=1$ we find that (\ref{6.15}) is exactly satisfied initially.
In addition the ratio between the surface area of a sphere
and the square of the radius
is in accord with that for the Friedmann-Lema\^\i tre
solution \cite{Peb} for expanding, curved space in the open, closed, and
``flat'' cases. This choice, together with (\ref{6.13}), defines
$\beta \;\forall \;\rho , \tau $.

The dynamics are given by (\ref{2.28}).  The slow speed limit of the
equation of motion is,
\begin{eqnarray}
\ddot r&=&\left \{ \left (\frac {\dot a}a\right )^2+ \frac {\displaystyle
2\left [
\frac {\dot a\ddot a}{a^2}-\left (\frac {\dot a}a\right )^3-\frac {\dot a
c^2}{a^3R_0^2}\right ]}{\displaystyle \left [ \left (\frac {\dot a}a\right )^2
+\frac {c^2}{(aR_0)^2}+\frac {2GM}{r^3}-\frac {c^2}{r^2}\left [1-f^2(\rho )
\right ] \right ]^{1/2}}\right \}r \nonumber \\
&&-\frac {GM}{r^2}+\frac {c^2C_0^2}{r^3} .
\label{6.18} \end{eqnarray}
The curvature of space and the expansion of the universe contribute corrections
of second order which are currently undetectable in the planetary motions of
our solar system.

The solutions of this section have assumed that:
(i) There is spherical symmetry about the mass-energy concentration.
(ii) $T^4_1=T_4^1=0$.
(iii) $T^4_4$ is homogeneous as specified by (\ref {6.11}),
but may be time dependent.
(iv) The metric matches in the short distance limit, the form of the
Lema\^\i tre
form of the static Schwarzschild metric corresponding to a mass-energy
concentration of strength $M$. That is to say, the region in which the
gravitational force is much greater than that of the expansion of the universe.
 (v) The metric matches the long distance form of the
Friedmann-Lema\^\i tre metric. Under these assumptions, the solutions of this
section is substantially unique, subject to the usual freedom involving
changes of variables. This result is a parallel to Birkhoff's theorem\cite{Bir},
but involves a different class of models than the class he considered.

\section {PROPERTIES OF THE HOMOGENEOUS, MASS-ENERGY DENSITY SOLUTIONS}

It is convenient to make the change of variables,
\begin{equation}
e^{\beta (\rho , \tau)}=a(\tau )e^{\sigma (\rho ,\tau)}. \label {7.1}
\end{equation}
Then (\ref {6.13}) becomes,
\begin{equation}
\dot \sigma ^2+2\frac {\dot a}a\dot \sigma= \frac {c^2}{(aR_0)^2}+
2GMe^{-3\beta } -c^2e^{-2\beta }\left [ 1-f^2(\rho )\right ] ,\label {7.2}
\end{equation}
which may be solved by means of the quadratic formula to give
\begin{equation}
\dot \sigma =\frac {\dot a}a\pm\left [\left (\frac {\dot a}a\right )^2+
\frac {c^2}{(aR_0)^2}+ \frac {2GM}{e^{3\beta }} -\frac {c^2\hat \rho ^2}
{R_0^2e^{2\beta }}\right ]^{1/2}, \label{7.3}
\end{equation}
where I have chosen a simple reparameterization of
(\ref{6.12}-\ref {6.17}), to wit,
\begin{equation}
\hat \rho =R_0\sin \left (\frac \rho {R_0}\right ) \label {7.4}
\end{equation}
so that (\ref{6.17}) becomes,
\begin{equation}
f(\hat \rho )=\sqrt{1-\hat \rho ^2/R^2_0} \label {7.5}
\end{equation}
I choose as initial conditions for (\ref{7.3}),
\begin{equation}
a(\tau _0)=1.0,\qquad \exp \left [\sigma (\hat \rho ,\tau _0)\right ]
= \hat \rho \label {7.5a}
\end{equation}

We may now check the special case, $M=0$ and $R_0= \infty $.  As we
expect $e^\beta = a(\tau )\hat \rho $, it follows that $\dot \sigma = 0$
for this case. That result in turn implies the minus sign in (\ref {7.3})
as $\dot a$ is known to be positive.
Next we check the case where $e^\beta $ is very small.  The dominant terms
in (\ref{7.3}) are,
\begin{equation}
\dot \sigma = \pm \sqrt{\frac {2GM}{(a(\tau )e^\sigma )^3}}.\label {7.6}
\end{equation}
Integrating this equation with respect to $\tau $ yields,
\begin{equation}
\exp \left [{\textstyle \frac 32\sigma (\hat \rho ,\tau )}\right ]-
\exp \left [{\textstyle \frac 32\sigma (\hat \rho ,\tau _0)}\right ]
=\pm \frac 32\sqrt{\frac {2GM}{c^2}}c\int _{\tau _0}^\tau a^{-3/2}(t)\, dt
\label {7.7}
\end{equation}
When we note that the integral in (\ref{7.7}) is approximately $\tau -
\tau _0$, and by use of (\ref{7.5a}), we get,
\begin{equation}
\exp \left [{\textstyle \frac 32\sigma (\hat \rho ,\tau )}\right ]\approx
\hat \rho ^{3/2} \pm \frac 32\sqrt{\frac {2GM}{c^2}}c(\tau -\tau _0)
\label{7.8} \end{equation}
In order to match the Schwarzschild-Lema\^\i tre metric (\ref {2.1c})
for small $e^\beta $
we must choose the minus sign in (\ref{7.8}) and hence also in (\ref{7.3}).
From the examination of these special cases, we conclude in general (assuming
that $\dot a\geq 0$) that (\ref{7.3}) must always be taken as,
\begin{equation}
\dot {U} =\frac {3\dot a}{2a}{U}-\frac 32
\left [\left (\frac { \dot a}a{U}\right )^2+\frac {2GM}{a^3(\tau )}
+\frac {c^2\root 3 \of {{U}^2}}{(a(\tau )R_0)^2}\left (
\root 3\of {{U}^4} -\hat \rho ^2\right ) \right ]^{1/2}, \label{7.9}
\end{equation}
where ${U}\equiv \exp (\frac 32\sigma )$.

It is worth pointing out that
when $a\equiv 1.0$ that (\ref{7.9}) subject to the initial conditions
(\ref{7.5}-\ref{7.5a}) gives the {\it static, curved-space solution} to the
Einstein field equations for a mass concentration in a vacuum.
In this special case, $\beta (\rho , \tau )= \sigma (\rho , \tau )$ by
(\ref{7.1}).  Eq.\ (\ref{7.9}) becomes in this case,
\begin{equation}
\dot {U} =-\frac 32
\left [2GM
+\frac {c^2\root 3 \of {{U}^2}}{R_0^2}\left (
\root 3\of {{U}^4} -\hat \rho ^2\right ) \right ]^{1/2}, \label{7.9a}
\end{equation}
where here $U\equiv \exp (\frac 32\beta )$. The integration of this equation
follows directly using the aforementioned initial conditions. $\alpha
(\rho ,\tau )$ follows by (\ref{2.3}) and (\ref{7.5}).
By (\ref{6.11}) and (\ref{6.14}), we
conclude that the $ T^1_1=T^2_2=T^3_3=3/(8\pi R_0^2)$ and $8\pi T^4_4=R_0^{-2}$,
as expected in a curved-space vacuum.

The time when an observer at rest with respect to the coordinate system,
{\it i.e.}\ a freely falling observer, arrives at the mass concentration,
{\it i.e.}, when $e^\sigma = 0$ for a given $\hat \rho $ with initial
conditions (\ref{7.5a}), is given by
\begin{eqnarray}
&&{U}(\hat \rho,\tau _{\rm zero})-{U}(\hat \rho,\tau _0)
 =-\hat \rho ^{3/2} = \nonumber \\
&&-\frac 32\int _{\tau _0}^{\tau _{\rm zero }}dt\, \Bigg \{
\left [\frac {2GM}{a^3(\tau )}+\left (\frac { \dot a}a{U}\right )^2
+\frac {c^2\root 3\of{{U}^2}}{(a(\tau )R_0)^2}\left (
\root 3\of {{U}^4} -\hat \rho ^2\right ) \right ]^{1/2} \label{7.10} \\
&&-\frac {3\dot a}{2a}{U}\Bigg \}
\qquad \tau _{\rm zero}\approx \tau _0+\frac {2\hat \rho ^{3/2}}{3\sqrt {2GM}}
\nonumber \end{eqnarray}
by (\ref{7.8}) and (\ref{7.9}) Thus the physically allowed range of time is
$0\leq \tau \leq \tau _{\rm zero} (\hat \rho )$.

For numerical computations it is convenient to choose dimensionless quantities.
I select,
\begin{eqnarray}
{\mathcal T}&=&H_0\tau ,\;\; {\mathcal U}=\left (\frac {H_0}c\right )^{3/2}U,
\;\; {\mathcal M}=\frac {GMH_0}{c^3},\;\;  \frac {\dot{\mathcal A}}
{\mathcal A} =\frac {\dot a}{H_0a},\;\; {\mathcal R}=\frac {H_0}c\hat \rho ,
\nonumber \\ {\bf R}_0&=&\frac {H_0}cR_0 . \label{7.11}
\end{eqnarray}
The initial conditions are
${\mathcal A(T}_0)=\dot{\mathcal A}({\mathcal T}_0)/{\mathcal A(T}_0)=1$,
${\mathcal U(R,T}_0)={\mathcal R}^{3/2}$, and ${\mathcal M}_\odot \approx
0.799\times 10^{-23} h_{50}$. For reference, one astronomical unit is about
$0.808\times 10^{-15}h_{50}$ in our dimensionless units and  a Megaparsec is
about $1.67\times 10^{-4}h_{50}$.  Thus we may rewrite
(\ref{7.9}) as
\begin{equation}
\dot {\mathcal U}=\frac {3\dot{\mathcal A}}{2{\mathcal A}}{\mathcal U}-\frac 32
\left [\left (\frac { \dot {\mathcal A}}{\mathcal A}{\mathcal U}\right )^2
+\frac {2{\mathcal M}}{{\mathcal A}^3({\mathcal T})}
+\frac {\root 3\of{{\mathcal U}^2}}{({\mathcal A({\mathcal T} )}{\bf R}_0)^2}
\left ( \root 3\of {{\mathcal U}^4} -{\mathcal R}^2\right ) \right ]^{1/2},
\label{7.12}
\end{equation}
I have integrated this equation by means of the Runge-Kutta method \cite{Pr}
adapted to double precision.

  There are several special cases for the solution of equation (\ref{7.12}).
First, when $e^\beta $ is large, the solution is just $e^\beta =\rho a(\tau )$,
both for the case of ``flat'' and curved space. This solution is just what one
expects for the case of expanding Friedmann-Lema\^\i tre expanding space.
The quantity $e^\alpha $ is given by (\ref{7.5}) and (\ref{2.3}) which
is, of course, different for ``flat'' and curved space.
I illustrate the large $e^\beta $ case, and the approach to it in fig.~2 for
the case of flat space and $a(\tau )= (\tau /\tau _0)^{2/3}$.  The initial
conditions (\ref{7.5a}) insure that $e^\beta/[\hat \rho a(\tau _0)]=1$ in
every case. The central mass has been chosen to be that of the Sun.

The next special case is when $e^\beta $ is small. In this case, in line
with with the Lema\^\i tre form of the Schwarzschild metric, it is
appropriate to plot the results vs.~$(\tau -\tau _0)/\hat \rho ^{3/2}$.
This is because this form of the (static) metric can be written as
\begin{equation}
\frac {e^\beta}{\hat \rho }= \left [\frac 32\sqrt{\frac {2GM}{c^2}}\left (
1-\frac {c(\tau -\tau _0)}{\hat \rho ^{3/2}}\right )\right ]^{2/3} \label {7.13}
\end{equation}
I illustrate this case and the the approach to this limit in fig.~3.  The
large dots indicate the expected $[\hat \rho - c(\tau -\tau _0)]^{2/3}$ limiting
curve with vertex at $\tau _{\rm zero}$.

The divergence at $\tau =0$ noted in fig.~2, leads to the question of the
nature of this divergence.  It turns out that it depends on the exponent
$\gamma $ defined in the limit as $\tau \to 0 $ by $a(\tau )\asymp
W\tau ^\gamma $.  By considering the dominant terms in (\ref{7.12}) for flat
space, we obtain the results,
\begin{eqnarray}
e^\beta &\asymp &\frac c{H_0}\left [\frac {2{\mathcal M}}{4(\gamma -\frac 13)^2
-\gamma ^2}\right ]^{1/3}\left ( \frac \tau {\tau _0}\right )^{2/3},\qquad
\qquad \gamma > \frac 23 , \label {7.14} \\
e^\beta &\asymp &\frac c{H_0}\left (\frac {\tau}{\tau _0}\right )^\gamma \left [
C_4^2(\rho )-\frac {3{\mathcal M}}{\gamma (2-3\gamma )}\tau ^{2-3\gamma }
\right ]^{1/3} , \qquad 0<\gamma < \frac 23 , \label {7.15} \\
e^\beta &\asymp &\frac c{H_0}\left [- \frac 92{\mathcal M}\left (\frac \tau
{\tau _0}\right )^2\ln \left (\frac \tau {\tau _0}\right )\right ]^{1/3} ,
\qquad \qquad \gamma = \frac 23 . \label{7.16}
\end{eqnarray}

The behavior for values of $\gamma $ not illustrated is qualitatively similar,
but differs, of course, in detail.

The effects of the curvature of space on $\beta $ is not very significant
either for small values of $e^\beta $ or for large values, as in these cases
the limiting values discussed above are obtained.  The effects however are
of some significance for intermediate values of $e^\beta $.  I have illustrated
an example in fig.~4.

\section{DOMAINS OF ATTRACTION}

In this section we will focus our attention on the slow-speed (\ref{6.18}),
flat-space\cite {deB,Hu} limit of (\ref{2.28}) for radial motion
only. That is to say, $C_0=0$. Hence, using the dimensionless variables of
(\ref{7.11}), we obtain,
\begin{equation}
\ddot {\mathfrak{R}} = \left (\frac {\dot {\mathcal A}}{\mathcal A}\right )^2
{\mathfrak{R}} +\frac {\displaystyle \left [
\frac {\dot {\mathcal A}\ddot {\mathcal A}}{{\mathcal A}^2}
-\left (\frac {\dot {\mathcal A}}{\mathcal A}\right )^3\right] {\mathfrak{R}}}
{\displaystyle \left [\left (\frac {\dot {\mathcal A}}{\mathcal A}\right )^2
+\frac {2{\mathcal M}}{{\mathfrak{R}}^3}
\right ]^{1/2}}-\frac {\mathcal M}{{\mathfrak{R}}^2}\label{8.1}
\end{equation}
where ${\mathfrak{R}}=H_0e^\beta /c$. First consider the case where the rate of
expansion of the universe is decreasing.  Then $\ddot {\mathcal A}<0$.  If
we divide (\ref{8.1}) by $(\dot {\mathcal A}/{\mathcal A})^2{\mathfrak{R}}$
and let $x={\mathcal MA}^2/({\mathfrak{R}}^3\dot {\mathcal A}^2
)$, then it becomes
\begin{equation}
\left (\frac {\mathcal A}{\dot {\mathcal A}}\right )^2\frac {\ddot {\mathfrak{R}}}
{\mathfrak{R}}=1-x-\frac 1{\sqrt{1+2x}} +\frac {\ddot {\mathcal AA}}{\dot {\mathcal
A}^2\sqrt {1+2x}} \label {8.2}
\end{equation}

Note that the $\ddot {\mathcal A}$ term in (\ref {8.2}) is uniformly
negative over
the allowed range of $x$, ($0\leq x\leq \infty $). Hence we may check all the
cases $\ddot {\mathcal A}\leq 0$ by setting it to zero.  This being done,
we may compute in a straight forward manner that for very small $x$, the
right-hand side is just $-1.5x^2$.  Differentiation with respect to $x$ shows
that the derivative
is uniformly negative.  Thus we conclude that if the rate of expansion
of the universe is not increasing, then according to the metric of found in
section IV, there is just one domain of attraction and it is centered on
the mass concentration as (\ref{8.2}) shows that if $\ddot {\mathcal A}\leq 0$,
then $\ddot {\mathfrak{R}} < 0$. The analogous computations for curved space
proceed in a similar manner, but are a bit more complicated.

Next let us consider the case where, as reported by Reiss
{\it et al.}\cite{Riess} and Perlmutter {\it et al.}\cite{Perl}, the
rate of expansion of the universe is increasing.  As for this case,
$\ddot {\mathcal A}>0$, it follows directly from (\ref{8.1}) that in
the limit as ${\mathfrak{R}}\to \infty $, $\ddot {\mathfrak{R}}>0$.  Since it
is also the case that when ${\mathfrak{R}}\to 0$, $\ddot {\mathfrak{R}}< 0$,
it follows that there exists an ${\mathfrak{R}}_D$ for which
$\ddot {\mathfrak{R}}=0$.  Therefore, we now have two domains of attraction.
One is $0\leq {\mathfrak{R}} < {\mathfrak{R}}_D$. In this domain, all test
particles at rest with respect to the coordinate system appear to be
accelerating towards the origin.  The other domain is ${\mathfrak{R}}_D<
{\mathfrak{R}}$. In this domain all test particles at rest with respect to
the coordinate system appear to be accelerating towards infinity. One expects
that over time, the neighborhood of the domain boundary will be swept clean of
particles.

Note is taken that the ``Swiss cheese model'' also has 2 domains of attraction.
This fact is independent of the sign of $\ddot a$.  Also the location of
the boundary described by ${\mathfrak{R}}$ increases with time.

At this point it is convenient to define the standard
acceleration/de\-cel\-er\-a\-tion parameter $q_0\equiv -\ddot {\mathcal AA}/
(\dot {\mathcal A}^2)$.  Then, setting $\ddot {\mathfrak{R}}= 0$, the equation
for the domain boundary becomes, by (\ref{8.2}),
\begin{equation}
0=1-x-\frac {q_0+1}{\sqrt{1+2x}} \label{8.3}
\end{equation}
The value of $q_0$ is given\cite{CPT} by the equation,
\begin{equation}
q_0={\textstyle \frac 12}\Omega _M-\Omega _\Lambda \label {8.4}
\end{equation}
where, combining the results of \cite{deB,Hu,Riess,Perl} we deduce the
values $\Omega _M=8\pi G\rho _0/(3H^2_0)\approx 0.19$ and $\Omega _\Lambda
=\Lambda c^2/(3H^2_0)\approx 0.81 $. These values yield $q_0\approx -0.715$.
Note that for ``flat'' space, $-1\leq q_0$, as $0\leq \Omega _M $.
Some simple manipulations convert (\ref{8.3}) into a cubic equation in $x$.
\begin{equation}
2x_D^3-3x_D^2+1=(1+q_0)^2 \label {8.5}
\end{equation}
This equation is simply solved by standard methods.  Its solutions, of
course, also include those for the other sign of the square root in (\ref{8.3}).
Evidently, the solutions of (\ref{8.3}) all lie in the range $0\leq x \leq 1$.
If $q_0\geq 0$, there
are no positive real roots of (\ref{8.3}) and we take $x_D(q_0)=0$.
If $q_0<0$, then there is one positive real solution $x_D(q_0)$ of (\ref{8.3}).
For the above mentioned case, $x_D(-0.715)\approx 0.825$.

We now have the {\it domain-size--mass relation},
\begin{equation}
r_D=\left [\frac {GMa^2}{x_D(q_0)\dot a^2}\right ]^{\frac 13} . \label {8.6}
\end{equation}
Here $r_D$ is the radius of the domain and $M$ is the interior mass.
We appeal to the uniqueness (in our class of models) reported at the
end of the fourth section to allow us to treat all the mass as concentrated
in the center of the domain. See also the result (\ref{2.12}) in this regard.
The {\it domain-size--mass relation}
is an important conclusion which is of particular interest if the rate
of expansion of the universe is accelerating, as has recently been
reported\cite{Riess,Perl}.

It is to be noted that many mass concentrations are rotating and so are
only axisymmetric, rather than spherically symmetric.  This means that
near the mass concentration the metric has some of the characteristics
of the Kerr metric. In terms of Cartesian coordinates, the differences in
the far-field between the Kerr metric and the Schwarzschild metric \cite{Steph}
are $g_{4i}\asymp 2G(\vec r\times \vec I\cdot \vec e_i) /(rc)^3$, where
the $\vec e_i$ are unit vectors in the $x,\; y,\; z$ directions and $\vec I$
is the angular momentum of the central mass concentration. In regions where
this correction is relatively small compared to the gravitation term
$2GM/(c^2r)$, {\it i.e.}, $|\vec r\times \vec I|/(Mcr^2)\ll 1$, we may
safely neglect this effect.  On the other hand, where the correction is
significant, the precise details of this effect will require further study.
In addition, our lack of a clear knowledge of the distribution of dark matter
also creates uncertainty concerning the details the above obtained results.

We now consider a few applications of the {\it domain-size--mass relation}.
If we use the solar mass, we get the radius of the domain
boundary to be about $417h_{50}^{-2/3}$ light years.  However, in this case
the density of stars is sufficiently high so as to perturb significantly this
result. The Oort cloud only extends out to about 0.8 light years. We remark
that the {\it domain-size--mass relation} only provides an upper limit on
the size of a structure that a given mass can be expected to hold together.
That there are smaller structures can presumably be explained on the basis
of other circumstances.

Next consider a globular cluster.  They are observed to have about
10,000 to 1,000,000 stars and to have diameters of several 10's of light
years to about 200 light years.  The computed domain boundary diameters for
systems of this size are about $9000h_{50}^{-2/3}$ - $42000h_{50}^{-2/3}$ light
years. This result is of
the substantially larger than that which is observed.  It may be
that they are reduced in size when they orbit near the galactic center.

Let us now consider our galaxy.  It is estimated to contain something like
200 billion stars and to have a mass of 700 billion to 1 trillion solar
masses. This mass implies a domain radius of $3.7h_{50}^{-2/3}$ to
$4.2h_{50}^{-2/3}$ million light years,
which way off from the observationally estimated diameter of about
100 thousand light years.

When we consider the Local Group and the Virgo Cluster, the problem becomes
more interesting.  Specifically, it has been estimated that the roughly
30 galaxies in the local group have a mass of about $3\times 10^{12}M_\odot $,
not counting intergalactic dark matter. This mass corresponds to about a
radius of $6h_{50}^{-2/3}$ million light years.  The observed value is roughly
4 million light years.  In the Virgo Cluster of around 2000 galaxies, the mass
is thought to be of the order of $10^{15}M_\odot $. This mass includes a
significant amount of dark matter and gives a radius of about $42h_{50}^{-2/3}$
million light years, compared to the observation that the Milky Way galaxy
lies about 65 million light years from center of the cluster. Considering
the errors and the fact that the Virgo Cluster appears not to be in equilibrium
yet, this agreement is not too bad.

In conclusion, it appears that for larger scale structures composed of
galaxies and inter-galactic space, the observed increase in the rate of
expansion may be an important feature in determining the size of
self-bound gravitating systems.  For smaller structures like galaxies,
globular clusters {\it etc.}\ other mechanisms are presumably dominant.

\begin{acknowledgements}
The author is pleased to acknowledge helpful conversations with N. Balazs,
S. Habib, P. O. Mazur, E. Motolla, and M. M. Nieto.
\end{acknowledgements}

\vfil \eject
\begin{figure}
\epsfig{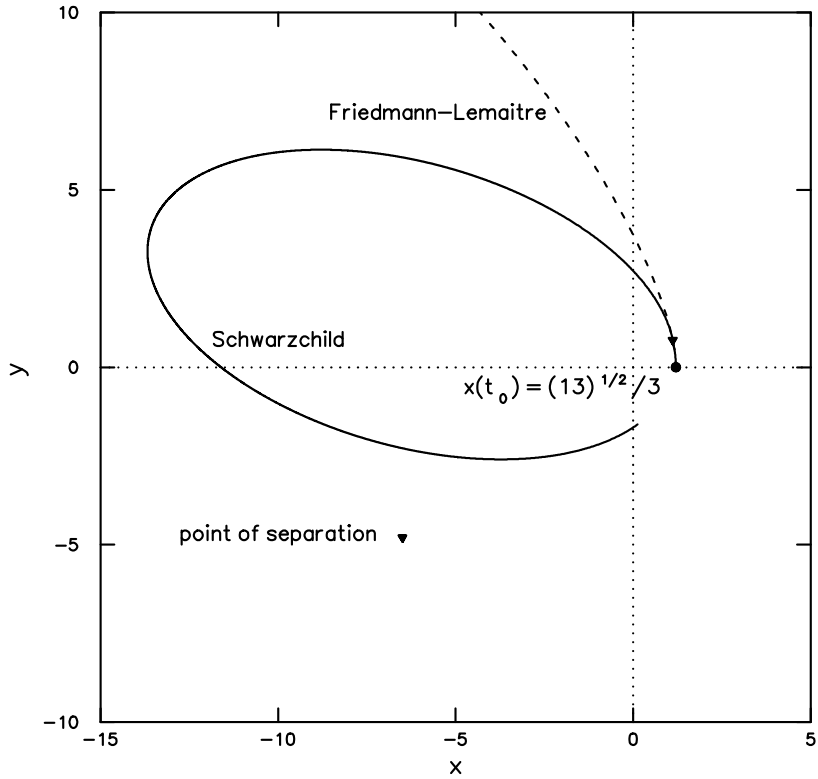}
\caption
{``Swiss cheese model'' trajectories which begin at
$x_0=\sqrt{13}/3 +\epsilon $ (the Friedman-Lema\^\i tre case), and
begin at $x_0=\sqrt{13}/3 -\epsilon $ (the Schwarzschild case). Here,
$\epsilon >0$ may be chosen as small as one pleases.  The initial part
of both trajectories is a parabola generated by the Friedmann-Lema\^\i tre
metric.  At time $t = (13/11)^3$, marked in the figure, the two trajectories
separate.}\label{fig.0}
\end{figure}

\vfil \eject
\begin{figure}
\epsfig{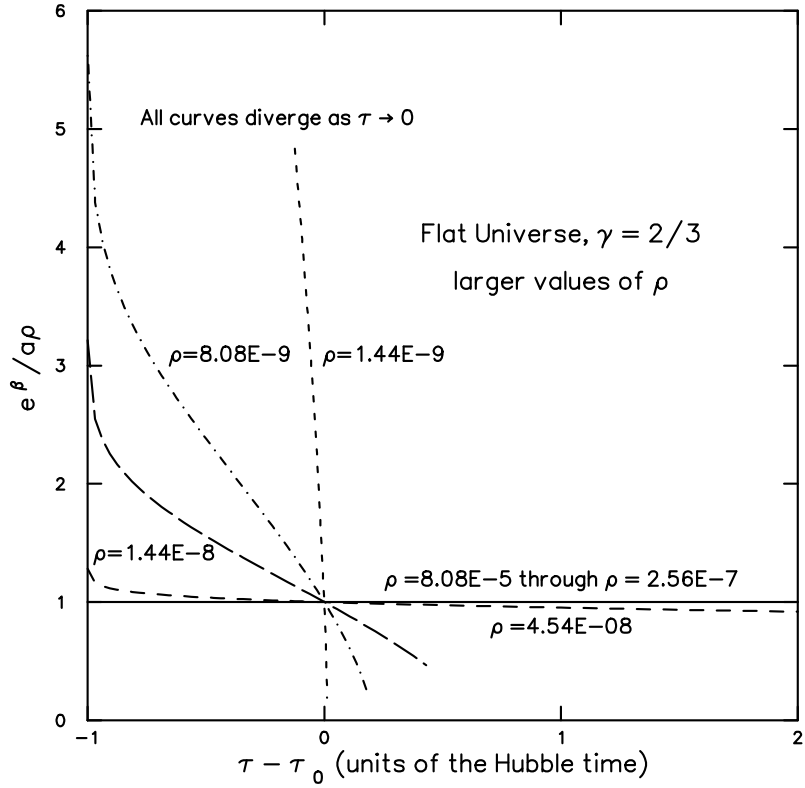}
\caption
{The behavior of $e^{\beta (\rho ,\tau )}$ for larger
values of $\rho $.  The solid line is the limiting value obtained for
large values of $\rho $. It is $e^\beta =a(\tau )\rho $.  The other lines
show how the limit is approached as $\rho $ increases.} \label{fig.1}
\end{figure}
\vfil \eject

\begin{figure}
\epsfig{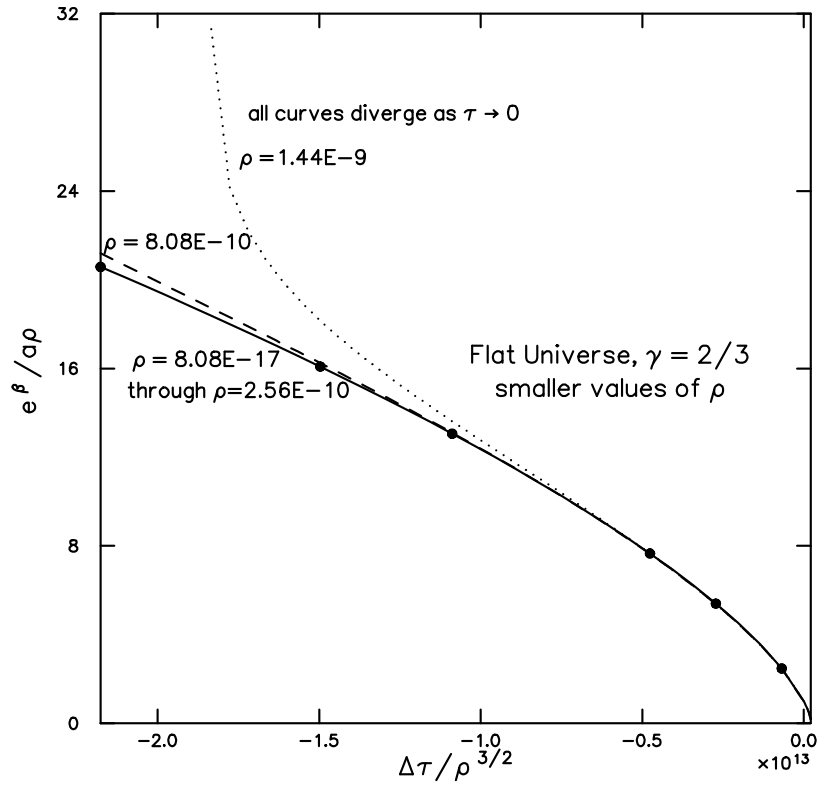}
\caption
{The behavior of $e^{\beta (\rho ,\tau )}$ for smaller
values of $\rho $.  The solid line is the limiting value obtained for
smaller values of $\rho $. It is $e^\beta \propto a(\tau )\rho [1-c
(\tau -\tau _{\rm zero})/\rho ^{3/2}]^{2/3}$.  The large dots show this
formula. The other lines show how the limit is approached as $\rho $
decreases.} \label{fig.2}
\end{figure}
\vfil \eject

\begin{figure}
\epsfig{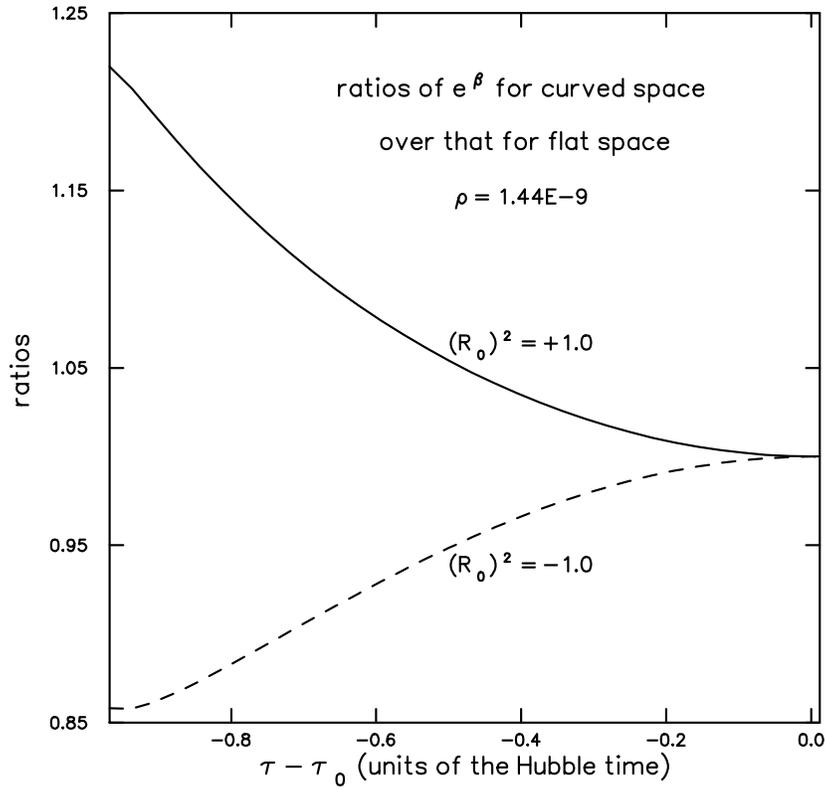}
\caption
{The variation of $e^{\beta (\rho ,\tau )}$ with the
curvature of space.  Displayed are the ratios of $e^\beta $ for curved
space to that for ``flat'' space with the same values of $\rho , \tau $.
The sold curve is for $R_0^2=+1$ and the dashed curve is for $R_0^2 =-1$.}
\label{fig.3}
\end{figure}
\end{article}
\end{document}